\documentstyle[preprint,tighten,floats,prd,aps]{revtex}
\input epsf

\newcommand{\be}{\begin{eqnarray}}
\newcommand{\ee}{\end{eqnarray}}
\begin{document}

\draft

\preprint{arch-ive/9611355 \hspace{104mm} UAHEP9611}

\title{
Gluino Contribution to the 3-loop $\beta$ function in the \linebreak
Minimal Supersymmetric Standard Model}
\author{L.\ Clavelli, P.\ W.\ Coulter and L.\ R.\ Surguladze}
\address{Department of Physics \& Astronomy, University of Alabama,
                   Tuscaloosa, AL 35487, USA}
\date{November 18, 1996}
\maketitle
\begin{abstract}
We deduce the gluino contribution to the three-loop QCD $\beta$ function
within the minimal supersymmetric Standard Model (MSSM) from its
standard QCD expression. The result is a first step in the computation of
the full MSSM three-loop $\beta$ function. In addition,
in the case of a light gluino it provides the strong three-loop SUSY
correction to the extrapolation of the strong coupling constant from
the low energy regime to the Z region and up to the squark threshold.
\end{abstract}


Although experimental measurements at the highest available energy are
consistent with the standard model \cite{exp}, the observed
relationship of the strong coupling constant at the Z and the weak
angle as well as the value of the $b/\tau$ mass ratio vis-a-vis the
top quark mass remain strong indications of a supersymmetric (SUSY)
grand unification above $10^{16} GeV$ and a SUSY threshold for squarks
and sleptons in the 0.1 to 1 TeV region. In this unification
picture the value of the SUSY threshold is very sensitive to the
highest known (two-loop) contribution to the MSSM $\beta$ function. At
the one-loop order a SUSY threshold far below 100 GeV would be needed
to fit the coupling constant measurements and such a low threshold is
directly ruled out by the non-observation of squarks and sleptons in Z
decay. This can be seen by inserting the current experimental values
for the couplings into the semi-analytic expressions for the SUSY
scale given in \cite{LC92}. This suggests that the three-loop results
could also be important especially as the precision of the
measurements at the Z and beyond improves. As a first step in the
calculation of the full three-loop $\beta$ functions of the MSSM, we
provide here the gluino contribution to the renormalization of the
strong coupling constant. This gives the complete result in the region
between the gluino mass and the squark mass which, in the light gluino
scenario, extends from the low energy regime up to the Z and beyond up
to the squark threshold. The standard Lagrangian density of the MSSM
is summarized in
\cite{Rosiek}.  The gluons interact with quarks and squarks in the
fundamental representation of the gauge group SU(N), and with gluons,
gluinos, and ghosts in the adjoint representation. In each
representation the generators satisfy the commutation relations
\be
         [R^a,R^b] = if^{abc}R^c  ,               \label{eq1}
\ee
with the adjoint representation matrices being defined in terms of the
structure constants $(F^a)_{bc} = if^{bac}$. The running of the strong
coupling constant as a function of the scale $\mu$ is determined by
the QCD $\beta$ function
\be
    d\alpha_s/d(\ln \mu^2) =  \alpha_s \beta(\alpha_s/{4\pi}),
\ee
where $\beta$ has the perturbative expansion
\be
    \beta(x) =  \beta_1 x + \beta_2 x^2+ \beta_3 x^3 + \ldots   .
\ee
Ignoring squark contributions,
the one- and two-loop results in the minimally extended SUSY QCD
\cite{Jones75} are
\be
 \beta_1 = -{11 \over 3} C_A + {4 \over 3}
  \biggl(n_f T + \frac{n_{\tilde{g}}}{2}C_A\biggr),
\ee
\be
 \beta_2 = - \frac{34}{3}C_A^2
   +\frac{20}{3}\biggl(n_f TC_A + \frac{n_{\tilde{g}}}{2}C_A^2\biggr)   
   +4\biggl(n_f TC_F + \frac{n_{\tilde{g}}}{2}C_A^2\biggr),    
\ee
where $n_f$ is the number of quark flavors and $n_{\tilde{g}}$ is the
number of gluino multiplets, $C_A$ and $C_F$ are the
eigenvalues of the quadratic
Casimir operators in the adjoint and fundamental representation
respectively and $T$ is the Dynkin index for the fundamental
representation. Note that, throughout this paper we use the framework
of the dimensional regularization \cite{drg} and a minimal subtraction
(MS, $\overline{\mbox{MS}}$) type prescriptions \cite{MSB}.
In the Standard Model ($n_{\tilde{g}}=0$), the three-loop coefficient is
\cite{TVZ}
\be
\beta_3^{\mbox{\scriptsize (SM)}} =
                 - \frac{2857}{54} C_A^3 - n_fT(2C_F^2
         - \frac{205}{9} C_F C_A - \frac{1415}{27} C_A^2)
   -(n_fT)^2 (\frac{44}{9} C_F+\frac{158}{27} C_A).       
\label{beta3sm}
\ee
Arriving at this result required the calculation of several hundred
Feynman graphs, and a similar number is required to extend the result
to the MSSM. However the first step in this program, the incorporation
of gluino loops, can be obtained by purely group theoretical methods.
This leaves a sharply reduced number of graphs that must be treated in
detail to determine the full three loop $\beta$ function. The main result
of this paper, proven in the appendix, is that, in the MSSM above the
gluino mass scale but below that of the squarks, there are only six
linearly independent group factors, $C_i,
\hskip 0.1in i=0,\ldots,5$, among all the relevant Feynman graphs.
In terms of these the three-loop $\beta$ function of the MSSM takes
the form:
\be
 \beta_3 = b_0 C_0 + b_1 C_1 + b_2 C_2 + b_3 C_3 + b_4 C_4 + b_5 C_5.
\label{beta3}
\ee
The graphs involving solely gluons and ghosts have each a group factor
proportional to $C_0 = C_A^3$ as can be trivially deduced on
dimensional grounds from the case where there are no quarks in the
theory. Graphs involving a single fermion loop have group weights that
are linear combinations of the three factors
\be
  C_1 = n_f T C_F^2 + n_{\tilde{g}} C_A^3/2   ,    \label{C1} 
\ee
\be
  C_2 = n_f T C_F C_A + n_{\tilde{g}} C_A^3/2    ,   \label{C2} 
\ee
\be
  C_3 = n_f T C_A^2 + n_{\tilde{g}} C_A^3/2.         \label{C3}
\ee
Finally, all graphs involving two fermion loops have group weights
that are linear combinations of the two factors
\be
 C_4 = (n_f T C_F + n_{\tilde{g}} C_A^2/2)(n_f T + n_{\tilde{g}} C_A/2),
\label{C4} 
\ee
\be
 C_5 = (n_f T C_A + n_{\tilde{g}} C_A^2/2)(n_f T + n_{\tilde{g}} C_A/2).
\label{C5}
\ee
In the standard model case where $n_{\tilde{g}} =0$, these results are trivial
and not useful. The importance of 
eq.\ (\ref{C1}) - (\ref{C5}) is that they
constrain the ways contributions from gluino loops follow from those
of quark loops. Substituting $C_0$ through $C_5$ into eq.\ (\ref{beta3}) 
and comparing with eq.\ (\ref{beta3sm}) in the $n_{\tilde{g}}=0$ limit suffices to
determine the coefficients $b_0$ through $b_5$. The final result for
the MSSM including gluinos but excluding squark contributions is
therefore
\begin{eqnarray}
\lefteqn{\hspace{-19mm}
   \beta_3 = - {2857 \over 54} C_A^3 - n_fT(2C_F^2
         - \frac{205}{9} C_F C_A - \frac{1415}{27} C_A^2)
             -(n_fT)^2 (\frac{44}{9} C_F+\frac{158}{27} C_A)}
                                                     \nonumber\\
 && \quad \hspace{15mm}
    + {988 \over 27}n_{\tilde{g}} C_A^3
     - n_{\tilde{g}} n_f T ({224 \over 27}C_A^2 +{22 \over 9} C_A C_F)
      - {145 \over 54}n_{\tilde{g}}^2 C_A^3.               
\label{final} 
\end{eqnarray}
Assuming the gluino lies below the Z and the squarks above the Z, then
at the Z scale, $\beta_3$ is $-9769/54$ in the standard model
($n_{\tilde{g}}=0$) and $+14134/27$ in the MSSM ($n_{\tilde{g}}=1$). If the gluino lies
below the $b$ quark, the value of $\alpha_s$ at the Z for a given
$\alpha_s$ at $m_b$ is increased by a non-negligible amount compared
to the experimental error. A complete analysis of the precise effect
is left for a later complete phenomenological analysis which should
include the light gluino effect on the Z and $\tau$ decay widths
\cite{CS,Chetyrkin}. The results found here for the MSSM
using the dimensional regularization \cite{drg} framework,
as expected,
do not match to those found recently \cite{JJN,FJJ} within the
dimensional reduction framework and they do not have to.
A similar scheme dependence has been found in \cite{TV}.

\acknowledgements

The authors acknowledge useful comments by Igor
Terekhov. This work was supported in part by the Department of Energy
under grant no. DE-FG02-96ER-40967.

\appendix

\section*{}

The group factors occuring in the $\beta$ function are calculated in
terms of traces of representation matrices. 
The fundamental
representation matrices, $T^a$, and the adjoint representation matrices,
$F^a$ satisfy
\be
T^aT^a = C_F \hat{\bf 1}, \hskip .5in Tr(T^aT^b)=T\delta^{ab},
\label{grpa}
\ee
\be
F^aF^a = C_A \hat{\bf 1}, \hskip .5in Tr(F^aF^b)=C_A \delta^{ab},
\label{grpb}
\ee
\be
F^aF^bF^a = C_A F^b /2,  \hskip .5in T^aT^bT^a = (C_F-C_A/2)T^b, \label{grpc}
\ee
\be
Tr(T^aT^bT^c) = T(d^{abc}+if^{abc})/2,  \hskip .5in
                                  Tr(F^aF^bF^c) = if^{abc} C_A /2,
\label{grpd}
\ee
\be
f^{abc}f^{abc} = C_A(N^2-1),
\label{grpe}
\ee
where
\be
 C_A = 2TN, \hskip .5in  C_F = T(N^2-1)/N.
\ee
The unit matrices appearing in eqs.\ (\ref{grpa}),(\ref{grpb}) are
$N$ and $N^2-1$ dimensional for the fundamental and adjoint
representations of SU(N)
respectively. The arbitrary normalization of the generators is usually
chosen so that $T=1/2$. The gauge coupling constant usually quoted in
experimental analyses uses this normalization. The result of this
work, that the three-loop $\beta$ function including gluino but not
squark contributions is a linear combination of the six quoted group
factors, is proven in this appendix.

We begin by noting that the MS scheme strong coupling renormalization
constant $Z_{\alpha_s}$ can be determined via 
the renormalization constants ${\tilde Z}_1$ for the
ghost-ghost-gluon vertex, $Z_3^{1/2}$ for the gluon propagator, and
${\tilde Z}_3^{1/2}$ for the ghost propagator (see, e.g., \cite{TV}
for more details)
\be
     Z_{\alpha_s} = {\tilde Z}_1^2 Z_3^{-1} {\tilde Z}_3^{-2}, 
\ee
where by the definition of the MS type schemes \cite{MSB}
each renormalization constant $Z_i$ is a polynomial
\begin{equation}
Z_i = 1 + \sum_{n\geq 1}Z^{(n)}_{i}(\alpha_s)\varepsilon^{-n},
\end{equation}
with $\varepsilon=(4-D)/2$. Now, the QCD $\beta$ function can be
obtained from the following relation (see, e.g.,\cite{TV})
\be
 \beta  =
        {{\partial Z^{(1)}_{\alpha_s}}\over{\partial \ln \alpha_s}}.
\ee

Contributions from graphs with no fermion loops each have the group
factor $C_A^3 = C_0$ as can be seen from the standard model result
eq.\ (\ref{beta3sm}). The relevant Feynman graphs involving at least one
fermion loop are those in fig.\ 1 with external gluons being
attached at all possible points. We may average over the $N^2-1$
color gluons. Thus the group factors that occur are $1/(N^2-1)$ times the
group factors of the graphs obtained by adding an internal gluon line
in all possible ways to the graphs of fig.\ 1, thus transforming
them into four-loop graphs. In each of these four-loop graphs there
are six gluon vertices. We may classify them by the resulting number
of gluons attached to the outer fermion loop in fig.\ 1a-d. We ignore
for the present quartic couplings. Each resulting graph will then have
n connections on the outer fermion line and 6-n connections in the
inner loop and we treat the graphs in order of decreasing n.
eq.\ (\ref{eq1}) 
tells us that the group factor for a graph with any order of gluon
attachments on the outer loop is a linear combination of the group
factor from the graph with the gluons attached in a standard order and
group factors from graphs with fewer gluon attachments on the outer
loop. This relation may be described by a graphical equivalence in
color space illustrated in fig.\ 2. Since eq.\ (\ref{eq1}) holds for all
representations, the same equivalence is valid on a gluon or ghost
loop also. Thus the linearly independent group factors may be found by
considering planar graphs only. When gluinos are incorporated into the
theory, each graph with a quark loop has a corresponding graph with
one or more quark loops replaced by gluino loops. The gluino
contribution to the $\beta$ function is given by the quark contribution
except for the replacement of the fundamental representation matrices
by the adjoint representation matrices and by an extra factor of 1/2
for each gluino loop due to the Majorana nature of the gluino. This
factor of 1/2 has its source in the fact that, unlike the case for
Dirac fermions, one must ignore the direction of the gluino line in a
Feynman graph in calculating statistical and symmetry factors.
The only linearly
independent group factor coming from graphs with n=6 may be taken as
the planar graph obtained by attaching another gluon line at adjacent
points on the outer loop of fig.\ 1c. Using the identities of
eqs.\ (\ref{grpa},\ref{grpb}),
the corresponding group factor is seen to be the $C_1$ of eq.\ (\ref{C1}).
\be
  C_1 = {1 \over {N^2-1}} \Bigl ( n_f \mbox{Tr}(T^aT^aT^bT^bT^cT^c)
        + {n_{\tilde{g}} \over 2} \mbox{Tr}(F^aF^aF^bF^bF^cF^c) \Bigr ).
\label{n6}
\ee
The traces here are easily evaluated using
eqs.\ (\ref{grpa})-(\ref{grpe}). Because of fig.\ 2, the group factor
from the various non-planar graphs with n=6 are linear combinations of
$C_1$ and group factors coming from graphs with $n < 6$. Thus
eq.\ (\ref{n6}) can be taken to be the only linearly independent group
factor
coming from the n=6 graphs. Graphs with n=5 can obtained by adding a
gluon line to fig.\ 1d or fig.\ 1c. Using fig.\ 2, one sees that all
such
graphs lead to group factors that are proportional to the $C_2$ of
eq.\ (\ref{C2}).
\be
C_2 = {-i \over {N^2-1}} \Bigl ( n_f \mbox{Tr}(T^aT^aT^bT^cT^d)
     + {n_{\tilde{g}} \over 2} \mbox{Tr}(F^aF^aF^bF^cF^d) \Bigr ) f^{bcd}
\label{n5}
\ee
plus possibly a linear combination of group factors from graphs with
lower n. It is clear, using the tracelessness of the representation
matrices and the equivalence of fig.\ 2, that this is the only linearly
independent group factor one can write with 5 attachments on the outer
loop and one internal attachment. We proceed, therefore to n=4. A
linearly independent group factor with n=4 is the $C_3$ of eq.\ (\ref{C3}).
\be
 C_3 = {4 \over {N^2-1}} \Bigl ( n_f \mbox{Tr}(T^aT^bT^cT^d)
    + {n_{\tilde{g}} \over 2} \mbox{Tr}(F^aF^bF^cF^d) \Bigr ) f^{abe}f^{cde}.
\label{n4}
\ee
This appears when the extra gluon line is added connecting the gluon
lines of fig.\ 1c. A second group invariant corresponding to n=4 is
\be
    {1 \over {N^2-1}} \Bigl ( n_f \mbox{Tr}(T^aT^aT^bT^c)
   + {n_{\tilde{g}} \over 2} \mbox{Tr}(F^aF^aF^bF^c) \Bigr ) \mbox{Tr}(F^bF^c).
\ee
However, by explicit calculation this is seen to be proportional to
$C_2$. At this point one should investigate the graph where to fig.\ 1d
one adds a gluon line with one leg on the outer loop and the other
attaching to make a quartic coupling. However the group factor at a
quartic coupling is a sum of three terms each of which is a product of
two $f^{abc}$. Thus in group space the quartic coupling is equivalent
to a sum of products of triple vertices as expressed in fig.\ 3. Again
this figure has the meaning that the group factor of a graph including
the quartic coupling is a linear combination of group factors where
the four legs are linked in the three possible ways by triple
couplings. The consequence of this equivalence is that we may neglect
all graphs with quartic couplings in determining the number and form
of linearly independent group factors. With n=4 we also have a
contribution from the two fermion loop topology of fig.\ 1a where a
gluon line is attached at adjacent points on the outer loop and each
of the fermion lines can represent either a quark or a gluino. The
corresponding group factor is $C_4$.
\be
C_4={1 \over{N^2-1}} \Bigl ( n_f \mbox{Tr} T^aT^aT^bT^c) + {n_{\tilde{g}} \over 2}
\mbox{Tr}(F^aF^aF^bF^c) \Bigr ) \Bigl ( n_f \mbox{Tr}(T^bT^c) +
{n_{\tilde{g}} \over 2} \mbox{Tr}(F^bF^c) \Bigr ).
\ee
All other graphs with four attachments on the outer loop can be seen,
using fig.\ 2, to be linear combinations of $C_1$ through $C_4$ and
possible group factors appearing at lower n. At n=3 we have the group
factor from fig.\ 1a with the extra gluon connecting the outer and
inner loops. The corresponding group factor is
\begin{eqnarray}
\lefteqn{ \hspace{-27mm}
     {{-2i}\over{N^2-1}} \Bigl ( n_f \mbox{Tr}(T^a[T^b,T^c]) +
     {n_{\tilde{g}} \over 2} \mbox{Tr}(F^a[F^b,F^c])\Bigr )}\nonumber\\
 && \quad \hspace{17mm}
        \Bigl ( n_f \mbox{Tr}(T^aT^bT^c)
   +{n_{\tilde{g}}\over 2} \mbox{Tr}(F^aF^bF^c) \Bigr )\nonumber\\
 && \quad \hspace{77mm}
                         = C_5. 
\end{eqnarray}
Here we have used the fact that the divergent part of fig.\ 1a when
gluon crossed graphs are included is totally antisymmetric in a,b,c.
This can be seen from the fact that the divergent sub-graph of fig.\ 1a
with an extra gluon joining the two fermion loops is a renormalization
of the triple gluon vertex. The other two-fermion-loop possibility
at n=3 is obtained by connecting the extra gluon line to fig.\ 1a with
one end on a triple gluon vertex. The corresponding group factor is
proportional to $C_5$ as can be seen by using fig.\ 2. The only
remaining possibility with n=3 is
\be
{1 \over{N^2-1}} \Bigl ( n_f \mbox{Tr}(T^a[T^b,T^c]) + {n_{\tilde{g}}/2}
 \mbox{Tr}(F^a[F^b,F^c] \Bigr ) \mbox{Tr}\Bigl ( F^aF^bF^c \Bigr ),
\ee
coming for example from adding a gluon line to fig.\ 1b or fig.\ 1d.
However this is easily seen to be proportional to $C_3$. Finally we
consider graphs with n=2. One such graph coming from fig.\ 1a is
equivalent to a graph treated earlier under interchange of the outer
and inner loops. The only other group invariants one can construct
with n=2 are of the form
\be
 C =  \frac{1}{N^2-1} \Bigl ( n_f \mbox{Tr} (T^aT^b) +
  {{n_{\tilde{g}}}\over 2} \mbox{Tr} (F^aF^b) \Bigr ) 
   \Bigl ( A \mbox{Tr} (F^aF^bF^cF^c)   
 + B \mbox{Tr}(F^aF^e) \mbox{Tr}(F^eF^b) \Bigr )  
 \ee
all of which are proportional to $C_3$. The $B$ term, in fact,
never appears in the three-loop $\beta$ function since after cutting a
gluon line in the four-loop graph to make the external gluons, the
graph must remain one-particle-irreducible. This exhausts the graphs
contributing to the gluon propagator renormalization.

The ghost
propagator renormalization contributes no new group factors. This can
be seen by noting that the ghost-ghost-gluon vertex has the same group
factor, $f^{abc}$, as the triple gluon vertex. The ghost propagator
corrections can be obtained from the topologies of fig.\ 1 by adding a
gluon in all possible ways which lead to at least one closed gluon
loop and then replacing each gluon loop in turn by a ghost loop.
Cutting one of the ghost lines then leads to a ghost propagator
renormalization graph. All of these clearly have the same group
structure as the corresponding contribution to the gluon propagator
renormalization. In fact, since the role of the ghosts is to cancel
unphysical Lorentz modes of the gluons, one can anticipate that the
ghost propagator and vertex renormalizations contribute no new
linearly independent group factors. It is useful, however, to see this
in detail.

We therefore consider finally contributions to the $\beta$
function from renormalization of the ghost-ghost-gluon coupling. The
corresponding graphs are given by attaching an external gluon in all
possible ways to the graphs of fig.\ 1 and then at all possible points
joined by a continuous gluon line attaching a pair of external ghosts
and changing the continuous gluon line between them into a ghost line.
Since the ghost-ghost-gluon vertex like the triple gluon vertex is
proportional to $f^{abc}$, the infinite corrections to the vertex must
also have this group structure. Since $f^{abc}f^{abc}=C_A(N^2-1)$, all
the group factors from the renormalization of this vertex can be
obtained by considering all the vacuum to vacuum, five loop graphs
including a fermion loop and a ghost loop that remain one-particle
irreducible after at least one particular gluon-ghost-ghost vertex is
excised. The group factors that occur in the gluon-ghost-ghost vertex
renormalization are then the group factors of these five-loop graphs
divided by $C_A(N^2-1)$. Each of these graphs consists of a ghost loop
and fermion loop with $3 < n < 6$ gluon vertices on the ghost
loop. Four representative graphs are shown in fig.\ 4. The three-loop
ghost-ghost-gluon vertex correction can be restored by excising a
ghost-ghost-gluon vertex from these five-loop graphs in all possible
ways that leave the graph one-particle irreducible. Since there are
eight gluon vertices in a five loop graph, all the graphs with six
attachments to the ghost loop will have two attachments to the fermion
loop as in fig.\ 4d. Each such graph, therefore, will have a factor
\be
 n_f \mbox{Tr} (T^aT^b) + \frac{n_{\tilde{g}}}{2} \mbox{Tr} (F^aF^b) =
 (n_f T + \frac{n_{\tilde{g}}}{2} C_A) \delta^{ab}.
\ee
The group factor for such graphs will therefore be
$(n_f T + n_{\tilde{g}} C_A/2)/C_A$ times the group factor $(C_A^3)$
of the three-loop gluon
propagator correction with no fermion loops. The resultant group
factor is $C_3$. A graph such as fig.\ 4a with three gluons attached to
the ghost loop clearly gives no new group factors since
\be
    \frac {1}{C_A(N^2-1)} \mbox{Tr} (F^gF^hF^i) =
    \frac{if^{ghi}}{2(N^2-1)}.
\ee
That is, a graph with a ghost loop attached to the rest of the graph
by three gluons has the same group factor as the graph where the ghost
loop is shrunk to a point at a triple gluon vertex leading to one of
the graphs treated in the gluon propagator renormalization. Graphs
with 5 vertices on the ghost loop (e.g, fig.\ 4b) have the group factors
\be
    \frac{1}{C_A(N^2-1)} \Bigl( n_f \mbox{Tr} (T^bT^cT^d)
    + \frac{n_{\tilde{g}}}{2} \mbox{Tr} (F^bF^cF^d) \Bigr )
    \mbox{Tr} (F^aF^aF^bF^cF^d)
\label{A9}
\ee
and
\be
    \frac{1}{C_A(N^2-1)} \Bigl( n_f \mbox{Tr}(T^bT^cT^d)
    + \frac{n_{\tilde{g}}}{2} \mbox{Tr} (F^bF^cF^d) \Bigr)
    \mbox{Tr} (F^aF^bF^aF^cF^d).
\label{A10}
\ee
Each of these has a group factor proportional to $C_3$ as can be seen
by using eqs.\ (\ref{grpa})-(\ref{grpc}). Graphs where there is a
triple
gluon vertex can be reduced to a linear combination of
eqs.\ (\ref{A9}) and (\ref{A10}) using the equivalence of
fig.\ 2. The graph of
fig.\ 4c where the fermion and ghost loop are joined by four gluons
has, when all the crossed gluon graphs are considered, the group
factor
\be
  \frac {1}{C_A(N^2-1)} \Bigl( n_f \mbox{Tr}(T^aT^bT^cT^d)
  + \frac{n_{\tilde{g}}}{2} \mbox{Tr} (F^aF^bF^cF^d) \Bigr) 
     \Bigl( A \mbox{Tr} ([F^a,F^b]F^cF^d)  \nonumber \\ 
  + B \mbox{Tr}(F^a[F^b,F^c]F^d) +
    C \mbox{Tr}(F^aF^bF^cF^d-F^cF^bF^aF^d) \Bigr).
\ee
The symmetry of the divergent part of these graphs is determined from
the fact that the divergent sub-graph is a renormalization of the
quartic gluon vertex. Each of these three terms is proportional to
$C_3$. The group factor for fig. 4f is also proportional to $C_3$ as
is that for the case of fig.\ 4f with fermion and gluon loops
interchanged. Finally there is also a two-fermion-loop graph
renormalizing the ghost-ghost-gluon vertex. This is shown in fig.\ 4e
where the vertex renormalization is obtained by excising and
discarding a ghost-ghost-gluon vertex that leaves a connection from
the ghost loop to each of the fermion loops. The group factor for
fig.\ 4e is
\begin{eqnarray}
\lefteqn{ \hspace{-17mm}
   \frac{1}{C_A(N^2-1)} \Bigl( n_f \mbox{Tr} (T^aT^bT^c) +
 \frac{n_{\tilde{g}}}{2} \mbox{Tr} (F^aF^bF^c) \Bigr)}\nonumber\\
 && \quad \hspace{23mm}
       \Bigl( n_f \mbox{Tr} (T^cT^d)  
  + \frac{n_{\tilde{g}}}{2} \mbox{Tr}(F^cF^d) \Bigr) \mbox{Tr}(F^aF^bF^d) 
              = -{{C_5}\over{4}}.  
\end{eqnarray}
This completes the proof that, when gluinos are included, there are
only six linearly independent group factors in the three-loop $\beta$
function for the SU(N) strong coupling constant, namely $C_A^3$ and the
five group factors given in eqs.\ (\ref{C1})-(\ref{C5}).
This leads to the result stated in
eqs.\ (\ref{beta3}) and (\ref{final}) of the text.

\newpage

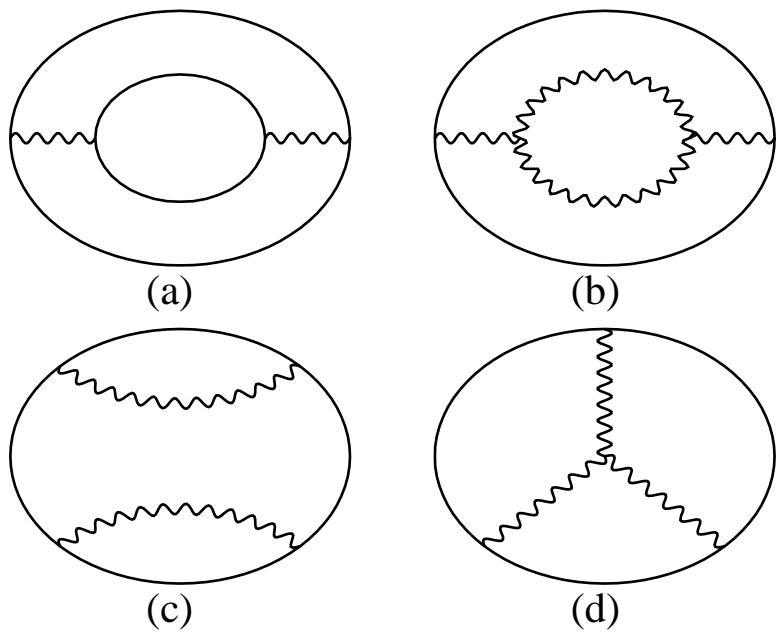
\begin{figure}
\caption{ Basic topologies for the calculation of the gluon propagator
renormalization. Solid lines represent fermions, either quarks or
gluinos, and wavy lines represent gluons. The pair of external gluons
are attached at all possible positions.}
\label{fig1}
\end{figure}

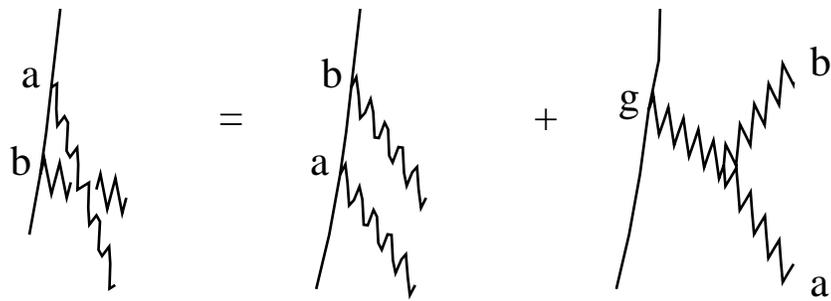
\begin{figure}
\caption{ Diagrammatic equivalence in group space.  The group factor
from a graph containing the left hand side is a linear combination of
the group factors from graphs in which each of the terms on the right
hand side is inserted in its place. }
\label{fig2}
\end{figure}

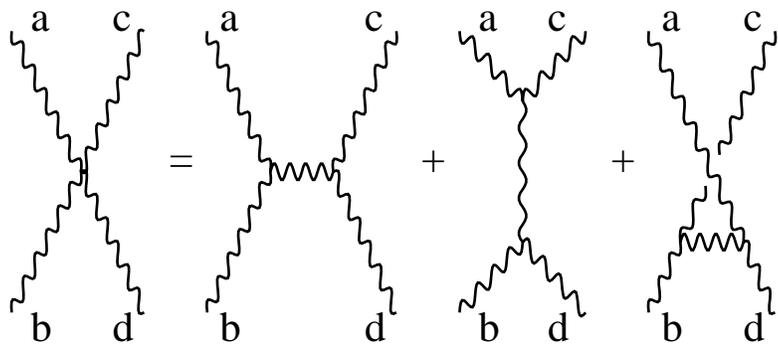
\begin{figure}
\caption{ Graphical equality in group space between a quartic coupling
and three pairs of triple couplings. As in fig.\ 2, the equality has
the meaning that the group factor of a graph containing the quartic
coupling is a linear combination of the group factors obtained by
replacing that coupling in turn by each of the subgraphs shown on the
right hand side.}
\label{fig3}
\end{figure}

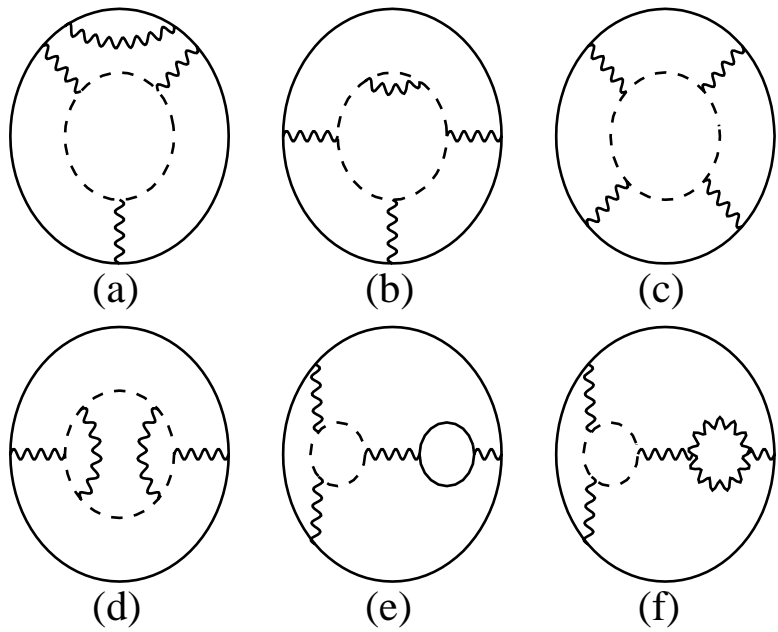
\begin{figure}
\caption{ Basic topologies with a least one fermion loop for the
renormalization of the ghost-ghost-gluon vertex. Fermions, gluons, and
ghosts are indicated by solid, wavy, and dashed lines respectively. }
\label{fig4}
\end{figure}

\end{document}